\documentclass[submission,copyright,creativecommons]{eptcs}
\usepackage{amsthm}
\usepackage{xcolor}
\usepackage{xspace}

\providecommand{\event}{ACL2 2015} 
\newtheorem{theorem}{Theorem}
\theoremstyle{definition}
\newtheorem{definition}{Definition}

\colorlet{light-gray}{gray!20}

\definecolor{dkgreen}{rgb}{0,0.4,0}
\definecolor{gray}{rgb}{0.4,0.4,0.4}
\definecolor{black}{rgb}{0.0,0.0,0.0}
\definecolor{mauve}{rgb}{0.58,0,0.82}
\definecolor{orange}{rgb}{0.7,0.30,0}
\definecolor{dkblue}{rgb}{0.0,0.0,0.4}
\definecolor{ltyellow}{rgb}{1.0,1.0,0.92}
\definecolor{red}{rgb}{0.4,0.0,0.0}
\definecolor{ltred}{rgb}{0.7,0.0,0.0}
\definecolor{pp}{rgb}{0.0,0.4,0.0}
\definecolor{linkblue}{rgb}{0.0,0.0,1.0}

\newcommand{\Ldef}[1]{{\ttfamily{#1}}}

\newcommand{\Lfn}[1]{{\ttfamily{#1}}}

\hypersetup{%
   colorlinks=false,
   urlbordercolor=linkblue,
   pdfborderstyle={/S/U/W 0.5}
}
\newcommand{\Xtopicname}[1]{\texttt{#1}}
\newcommand{\Xtopiclink}[2]{\href{http://www.cs.utexas.edu/users/moore/acl2/manuals/current/manual/?topic=#1}{\Xtopicname{#2}}}
\newcommand{\Xtextlink}[2]{\href{http://www.cs.utexas.edu/users/moore/acl2/manuals/current/manual/?topic=#1}{#2}}

\title{Fix Your Types} 
\author{Sol Swords \qquad Jared Davis
\institute{Centaur Technology, Inc.\\
7600-C N. Capital of Texas Hwy, Suite 300\\
Austin, TX 78731}
\email{\{sswords,jared\}@centtech.com}
}

\def\titlerunning{Fix Your Types}
\def\authorrunning{Sol Swords and Jared Davis}
\begin{document}
\maketitle

\begin{abstract}

When using existing ACL2 datatype frameworks, many theorems require
type hypotheses.  These hypotheses slow down the theorem prover, are
tedious to write, and are easy to forget.  We describe a principled
approach to types that provides strong type safety and execution
efficiency while avoiding type hypotheses, and we present a library
that automates this approach.  Using this approach, types help you
catch programming errors and then get out of the way of theorem
proving.


  
\end{abstract}



\section{Introduction}

ACL2 is often described as ``untyped,'' and this is certainly true to some
degree.  Terms like \texttt{(+ 0 "hello")}, which would be not be accepted by
the static type checks of languages like Java, are legal and well-defined in
the logic, and can even be executed so long as \Xtextlink{ACL2____GUARD}{guard}
checking is disabled.  Terms like \texttt{(/ 5 0)}, which would be well-typed
but would cause a run-time error in most programming languages, are also
logically well-defined and can also be executed when guards are not checked.

Of course, most ACL2 code is written with particular types in mind, often
expressed as the guards of the functions.  When proving properties of such
code, it's easy to get tripped up by corner cases where some variables of the
theorem are of the wrong types.  To avoid this, one strategy is to begin a
theorem with a list of type hypotheses, one for each variable mentioned.  These
hypotheses act as a kind of insurance: we may not know whether they're
necessary or not, but including them might save us from having to debug failed
proofs caused by missing type assumptions.

On the other hand, lists of type hypotheses are often repetitive, take time to
write, and make the formulas we're proving larger and less elegant.  We can
hide much of this with macros, e.g., the \Xtopiclink{ACL2____DEFINE}{define}
utility has special options like \texttt{:hyp :guard} for including the guard
as a hypothesis in return-value theorems.  But even then, these hypotheses will
cause extra work for the rewriter since it must relieve them before it can
apply our theorem, and may make it harder to carry out later proofs since our
theorem will not be applied unless its hypotheses can be relieved.

Accordingly, after we have proven a theorem, a good practice is to try to
``strengthen'' it by removing any unnecessary hypotheses.  There is even a
tool, \Xtopiclink{ACL2____REMOVE-HYPS}{remove-hyps}, that tries to
automatically identify unnecessary hypotheses ex post facto.  While
strengthening theorems is useful, it is limited.  For instance, we (of course)
cannot eliminate type hypotheses that are actually necessary for the formula to
be a theorem.  It can also be tedious, e.g., automation such as
\texttt{define}'s \texttt{:hyp :guard} does not provide a convenient way to
remove parts of the guard.  This is not a purely theoretical
concern; see for instance ACL2
\href{http://github.com/acl2/acl2/issues/167}{Issue 167}, a request for such
a feature.

An alternate strategy, which is well-known and certainly not novel, is to more
carefully code our functions so that they always treat ill-typed inputs
according to some particular \textit{fixing convention}.  By following this
approach, we can typically avoid the need for type hypotheses altogether.  Many
examples of this approach can be found throughout ACL2.  To name a few:

\begin{itemize}
\item Arithmetic functions treat non-numbers as 0; functions expecting a
  particular type of number (integer, natural) treat anything else as 0---e.g.,
  \texttt{zp}, \texttt{nth}, \texttt{logbitp}.
\item Functions expecting a string treat a non-string as
  \texttt{""}---e.g., \texttt{char}, \texttt{(coerce x 'list)}.
\item Any atom is treated as \texttt{nil} by \texttt{car}, \texttt{cdr},
  \texttt{endp}, etc.
\item The \Xtopiclink{ACL2____STD_F2OSETS}{std/osets}~\cite{04-davis-osets} library functions treat non-sets as \texttt{nil}.
\end{itemize}

Following this strategy typically takes a small bit of initial setup, e.g., to
agree upon and implement a fixing convention.  But once this convention is in
place, type hypotheses can be eliminated from many theorems.  For instance, we
have unconditional theorems such as $a + b = b + a$ without hypotheses about
$a$ and $b$ being numbers, and $X \subseteq X$ without hypotheses about $X$
being a set.  By eliminating type hypotheses, these theorems become easier to
read and write, and can be more efficiently and reliably used to simplifying
later proof goals.

Unfortunately, existing datatype definition frameworks for ACL2 don't provide
any easy way to follow the fixing strategy.  For example, consider the
available macros for introducing product types, like
\Xtopiclink{ACL2____DEFSTRUCTURE}{def\-structure}~\cite{97-brock-defstructure},
\Xtopiclink{ACL2____DEFDATA}{defdata}~\cite{14-harsh-defdata}, and
\Xtopiclink{STD____DEFAGGREGATE}{defaggregate}.  These macros define
constructors and accessor functions that do not support any particular
convention for dealing with ill-typed fields or products.  Consider a simple
\texttt{student} structure introduced by:

\begin{verbatim}
  (defaggregate student
    ((name stringp)
     (age  natp)))
\end{verbatim}

The constructor and accessors for \texttt{student} structures will have strong
guards that are useful for revealing programming errors in function
definitions.  However, in the logic, nothing prevents us from invoking the
\texttt{student} constructor on ill-typed arguments.  For instance, in cases
like:
\begin{verbatim}
  (make-student :name 6 :age "Calista")
\end{verbatim}
the constructor fails to produce a valid \texttt{student-p}.  The accessors
suffer from similar problems, for instance the following term is equal to
\texttt{6}, which is not a well-typed student name:
\begin{verbatim}
  (student->name (make-student :name 6 :age "Calista"))
\end{verbatim}
Consequently, reasoning about these structures almost always requires
type hypotheses.  Since the types defined by these frameworks are
often found at the lowest levels of ACL2 models, these hypotheses
percolate upwards, infecting the entire the code base!






In this paper, we address this problem with the following contributions:

\begin{itemize}
\item We present, in precise terms, a \textbf{fixtype discipline} for working
  with types in ACL2 (Section \ref{sec:discipline}).  This discipline allows
  efficient reasoning via avoiding type hypotheses, strong type checking via
  ACL2's guard mechanism, and preserves efficient execution via
  \Xtopiclink{ACL2____MBE}{mbe}.

\item Manually following the fixtype discipline would be tedious.
  Accordingly, we present a new library, \Xtopiclink{ACL2____FTY}{FTY}
  (short for ``fixtypes''), which provides automation for following
  the discipline (Section \ref{sec:FTY_library}). The FTY library
  contains tools that automate the introduction of new types and
  assist with creating functions that ``properly'' operate on those
  types.
\end{itemize}

While there is room for improvement (Section \ref{sec:future-work}), the
approach and automation that we present is practical and scales up to complex
modeling efforts.  We have successfully used FTY as the type system for two
large libraries: \Xtextlink{ACL2____VL}{VL}, which processes Verilog and
SystemVerilog source code; and \Xtextlink{ACL2____SV}{SV}, a hardware modeling
and analysis framework.  VL, in particular, involves a very complex hierarchy
of types.  For instance, it includes a 30-way mutually recursive datatype that
represents SystemVerilog expressions, types, and related syntactic constructs.

\section{The Fixtype Discipline}
\label{sec:discipline}

We begin, in this section, by describing in precise terms a \textit{fixtype
  discipline} for working with types in ACL2.  In our experience, following
this discipline is an effective way to obtain the benefits of strong type
checking while keeping types out of the way of theorem proving.

The basic philosophy behind the fixtype discipline is that all functions that
take inputs of a particular type should treat any inputs that are \textit{not}
of that type in a consistent way.  This can be done using \textit{fixing
  functions}.

\begin{definition}
\label{fixing_function_def}
A \textbf{fixing function} $\mathit{fix}$ for a (unary) type predicate
$\mathit{typep}$ is a (unary) function that (1) always produces an object of
that type, and (2) is the identity on any object of that type. That is, it
satisfies:
\[
\begin{array}{ll}
\textrm{(1)} & \forall x : \mathit{typep}(\mathit{fix}(x)) \\
\textrm{(2)} & \forall x : \mathit{typep}(x) \Rightarrow \mathit{fix}(x) = x \\
\end{array}
\]

\end{definition}

Given a fixing function, an easy way to ensure that some new definition treats
all of its inputs in a type-consistent way is to immediately apply the
appropriate fixing function to each input before proceeding with the main body
of the function.  For guard-verified functions, this preliminary fixing can be
done for free using \Xtopiclink{ACL2____MBE}{mbe}.  Alternatively, if all
occurrences of an input variable in the function's body occur in contexts that
are already type-consistent, then explicit fixing isn't necessary.




When a function follows this approach, the fixing functions become
``transparent'' to that function.  For instance, since \texttt{nth} properly
fixes its index argument to a natural number, the following holds:

\begin{verbatim}
 (defthm nth-of-nfix
   (equal (nth (nfix n) x)
          (nth n x)))
\end{verbatim}

Given any fixing function, we can define a corresponding
\Xtextlink{ACL2____EQUIVALENCE}{equivalence} relation.  For instance, for
naturals, we can define \texttt{nat-equiv} as equality up to \texttt{nfix}:

\begin{verbatim}
 (defun nat-equiv (x y)
   (equal (nfix x) (nfix y)))
\end{verbatim}

Functions that properly fix their arguments will satisfy a
\Xtextlink{ACL2____CONGRUENCE}{congruence} for this equivalence under equality:
that is, they produce \texttt{equal} results when given equivalent arguments.
For instance, for \texttt{nth}:

\begin{verbatim}
 (defthm nat-equiv-congruence-for-nth
    (implies (nat-equiv n m)
             (equal (nth n x)
                    (nth m x)))
   :rule-classes :congruence)
\end{verbatim}

We can now define our fixtype discipline.

\begin{definition}
  A function follows the \textbf{fixtype discipline} if, for each typed
  input, the type has a corresponding fixing function and equivalence
  relation, and the function produces equal results given
  type-equivalent inputs.
\end{definition}

A consequence of following the fixtype discipline is that theorems can
avoid type hypotheses.





\begin{theorem}
\label{nohyps_theorem}
  Let $\mathit{typep}$ be a type and let $\equiv$ be the equivalence relation
  induced by a fixing function for $\mathit{typep}$.  Let $C(x)$ be a
  conjecture satisfying the congruence
 \[ (x \equiv x') \Rightarrow \left( C\left(x\right) \Leftrightarrow C\left(x'\right) \right). \]
 Then $C(x)$ is a theorem if and only if
 $\mathit{typep}(x) \Rightarrow C(x)$
 is a theorem.

\end{theorem}

This congruence means that $C(x)$ is a formula where the variable $x$
is consistently treated according to the fixing discipline for
$\mathit{typep}$; in this case, it isn't necessary to include
$\mathit{typep}(x)$ as a hypothesis.  This generalizes easily to
additional variables.










\section{The FTY Library}
\label{sec:FTY_library}

If we want to follow the fixtype discipline, all of our type predicates need to
have corresponding fixing functions and equivalence relations.  Also, as we
introduce new functions that operate on these types, we need to prove that
these functions satisfy the appropriate congruences, i.e., that they treat
their inputs in a type-consistent way.

Although these definitions and proofs are straightforward, it would be very
tedious to carry them out manually.  It would also be difficult to make use of
libraries like \texttt{std/util} or \texttt{defdata} since the functions these
frameworks introduce do not follow the fixtype discipline.  This is unfortunate
because these libraries really make it far more convenient to introduce new
types.

To address this, we have developed a new library, named FTY, which provides
several utilities to automate this boilerplate work and to facilitate the
introduction of new types that follow the discipline.  Among these utilities,
we have:

\begin{itemize}
\item \texttt{deffixtype}, which associates a type predicate with a fixing
  function and equivalence relation, for defining base types like \texttt{natp},
  \texttt{stringp}, and custom user-defined base types. (Section
  \ref{sec:deffixtype})
\item \texttt{deftypes} and associated utilities \texttt{defprod}, {\tt
  deftagsum}, \texttt{deflist}, \texttt{defalist}, and more, which define new
  derived fixtype-compliant product types, sum types, list types, etc. (Section
  \ref{sec:deftypes})
\item \texttt{deffixequiv} and \texttt{deffixequiv-mutual}, which prove the
  appropriate type congruences for functions that operate on these
  types. (Section \ref{sec:deffixequiv})
\end{itemize}

We now briefly describe these utilities.  We focus here on what these utilities
automate and how this helps to make it easier to follow the fixtype discipline.
More detailed information on how to practically make use of these utilities, their
available options, etc., can be found in the \Xtextlink{ACL2____FTY}{FTY
  documentation}~\cite{fty-xdoc} in the ACL2+Books Manual.

\subsection{Deffixtype}
\label{sec:deffixtype}

The FTY library uses an ACL2 \Xtopiclink{ACL2____TABLE}{table} to record the
associations between the name, predicate, fixing function, and equivalence
relation for each known type.  This information is used by many later FTY
utilities to improve automation.  For instance, when we define a new structure,
this table allows us to look up the right fixing function and equivalence
relation to use for each field just by its type, without needing to be
repetitively told the fixing function and equivalence relation for every field.

The \textbf{deffixtype} utility is used to register new \textit{base types},
i.e., types that are not defined in terms of other FTY types, with this table.
Here is an example, which registers a new type named \texttt{nat}, recognized
by \texttt{natp}, with fixing function \texttt{nfix}, and with the equivalence
relation \texttt{nat-equiv}:

\begin{verbatim}
 (deffixtype nat
   :pred  natp
   :fix   nfix
   :equiv nat-equiv)
\end{verbatim}

The type name does not need to be a function name; we typically use
the name of the predicate without the final ``p'' or ``-p.''  The
predicate and fixing function must always be provided by the user and
defined ahead of time.  
\texttt{Deffixtype} can automatically define the equivalence relation
based on the fixing function, or it can use an existing equivalence relation.

We usually do not need to invoke \texttt{deffixtype} directly.  FTY includes a
\Xtopiclink{FTY____BASETYPES}{basetypes} book that sets up these associations
for basic ACL2 types like naturals, integers, characters, Booleans, strings,
etc.  When new derived types are introduced by FTY macros like
\texttt{deftypes} (Section \ref{sec:deftypes}), they are automatically
registered with the table.  On the other hand, \texttt{deffixtype} is
occasionally useful for defining low-level custom base types, or types that use
special encodings, or that for some other reason we prefer not to introduce
with \texttt{deftypes}.

Choosing a good fixing function for a type is not always straightforward.  As
far as the fixtype discipline and the FTY library is concerned, any function
that satisfies Definition \ref{fixing_function_def} suffices.  However, the way
in which ill-typed objects are mapped into the type affects which functions
will have proper congruences for the induced equivalence relation.  Some
choices are dictated by pre-existing ACL2 conventions; for example, if we wrote
our own \texttt{my-nat-fix} function that coerced non-naturals to 5 instead of
0, then this fixing function wouldn't be transparent to built-in functions such
as \texttt{zp} and \texttt{nth}.





The fixing function's guard may optionally require that the input object
already be of the type.  This allows the fixing function to be coded so that it
is essentially free to execute, using \Xtopiclink{ACL2____MBE}{mbe} so that the
executable body is just the identity.  It is also generally useful to inline
the fixing function to avoid the small overhead of a function call.  For
example:

\begin{verbatim}
 (defun-inline string-fix (x)
   (declare (xargs :guard (stringp x)))
   (mbe :logic (if (stringp x) x "")
        :exec x))
\end{verbatim}

\subsection{Deftypes and Supporting Utilities}
\label{sec:deftypes}

Whereas \texttt{deffixtype} is useful for registering base types and special,
custom types, the \textbf{deftypes} suite of tools can be used to easily define
common kinds of derived types.  The constructors, accessors, and other
supporting functions introduced for these types follow the fixtype discipline,
and the new types are automatically registered with \texttt{deffixtype}.  There
are utilities for introducing many kinds of types:

\begin{itemize}
\item \texttt{defprod}, which defines a product type,
\item \texttt{deftagsum}, which defines a tagged sum of products,
\item \texttt{deflist}, which defines a list type which has elements of a given type,
\item \texttt{defalist}, which defines an alist type with keys and values of given types,
\item \texttt{defoption}, which defines an option/maybe type,
\item and others.
\end{itemize}

Using these macros is not much different than using other data definition
libraries.  For instance, we can introduce a basic \texttt{student} structure
as follows:

\begin{verbatim}
  (defprod student
    ((name stringp)
     (age  natp)))
\end{verbatim}

This is very much like introducting a structure with
\Xtopiclink{STD____DEFAGGREGATE}{defaggregate}: it produces a recognizer,
constructor, accessors for the fields, \Xtopiclink{ACL2____B_A2}{b*} binders,
and readable make/change macros.  Unlike \texttt{defaggregate}, it also
generates a fixing function, \texttt{student-fix}, an equivalence relation,
\texttt{student-equiv}, and registers the new student type with
\texttt{deffixtype}.  The constructor and accessor functions for the new type
also follow the fixtype discipline, e.g., we unconditionally have theorems such
as:

\begin{itemize}
\item \texttt{(student-p (student name age))}
\item \texttt{(stringp (student->name x))}
\item \texttt{(natp (student->age x))}
\end{itemize}

A notable feature of \texttt{deftypes} is that it also provides strong support
for mutually recursive types.  In particular, several calls of utilities such
as \texttt{defprod}, \texttt{deflist}, etc., may be combined inside a
\texttt{deftypes} form to create a mutually-recursive clique of types.  For
example, to model a simple arithmetic term language such as:

\[
\begin{array}{lrll}
\mathrm{aterm} & = & \textsf{Num}~\{\texttt{val} :: \mathrm{integer}\} \\
               &   | & \textsf{Sum}~\{\texttt{args} :: \mathrm{List~aterm}\} \\
               &   | & \textsf{Minus}~\{\texttt{arg} :: \mathrm{aterm}\}
\end{array}
\]

\noindent We might write the following \texttt{deftypes} form:


\begin{verbatim}
  (deftypes arithmetic-terms
    (deftagsum aterm
      (:num   ((val integerp)))
      (:sum   ((args atermlist)))
      (:minus ((arg aterm))))
    (deflist atermlist
      :elt-type aterm))
\end{verbatim}
As you might expect, this form creates the basic predicates, fixing functions,
and equivalence relations for \texttt{aterm}s that are needed for the fixtype
discipline:
\begin{itemize}
\item Predicates \texttt{aterm-p} and \texttt{atermlist-p},
\item Fixing functions \texttt{aterm-fix} and \texttt{atermlist-fix}, and
\item Equivalence relations \texttt{aterm-equiv} and \texttt{atermlist-equiv}.
\end{itemize}
It also registers the new types with \texttt{deffixtype}.  The form also
defines several functions and tools for working with these new types, all of
which have appropriate congruences for the fixtype discipline:

\begin{itemize}

\item A kind function, \texttt{aterm-kind}, to determine the kind of an
  \texttt{aterm}, e.g., \texttt{:num}, \texttt{:sum}, or \texttt{:minus}.

\item Constructors for each kind of \texttt{aterm}: \texttt{aterm-num},
  \texttt{aterm-sum}, and \texttt{aterm-minus}, and associated make/change
  macros in the style of \texttt{defaggregate}/\texttt{defprod}.

\item Accessors for each kind of \texttt{aterm}: \texttt{aterm-num->val},
  \texttt{aterm-sum->args}, \texttt{aterm-minus->arg} and associated
  \texttt{b*} binders.

\item Measure functions, \texttt{aterm-count} and {\tt atermlist-count},
  appropriate for structurally recurring over objects of these types.
\end{itemize}

For convenience, a macro \texttt{aterm-case} is also introduced.  This macro
allows us to implement the common coding scheme of cases on the kind of an
\texttt{aterm}, followed by binding variables to any needed fields of the
product.  Here is a simple example of using \texttt{aterm} structures.


\begin{verbatim}
 (defines aterm-eval
   (define aterm-eval ((x aterm-p))
     :measure (aterm-count x)
     :returns (val integerp)
     :verify-guards nil
     (aterm-case x
       :num x.val
       :sum (atermlist-sum x.args)
       :minus (- (aterm-eval x.arg))))

   (define atermlist-sum ((x atermlist-p))
     :measure (atermlist-count x)
     :returns (val integerp)
     (if (atom x)
         0
       (+ (aterm-eval (car x))
          (atermlist-sum (cdr x)))))
   ///
   (verify-guards aterm-eval))
\end{verbatim}


\subsection{Deffixequiv and Deffixequiv-mutual}
\label{sec:deffixequiv}


Together, \texttt{deffixtype} and \texttt{deftypes} allow us to largely
automate the process of introducing new types that support the fixtype
discipline.  But this is only half the battle.  When we define new functions
that make use of these types, we are still left with the task of proving that
these functions satisfy the appropriate congruences for every argument of these
types.  If our model or program involves many function definitions, this can be
a lot of tedious work.

To automate this process, FTY offers two related utilities,
\textbf{deffixequiv} and \textbf{deffixequiv-mutual}.  These utilities are
integrated with \Xtopiclink{ACL2____DEFINE}{define} and
\Xtopiclink{ACL2____DEFINES}{defines} and also make use of the table of types
from \texttt{deffixtype}.  This allows them to figure out what theorems are
needed, often without any help at all.  In particular, the types of the
arguments are inferred from the extended formals of the each function.
The corresponding fixing functions and equivalence relations can then be looked
up from the table, and the appropriate congruence rules can be generated.
Besides congruence rules, we additionally generate rules that
normalize constant arguments to their type-fixed forms.

Consider the \texttt{aterm-eval} example above.  To generate the congruence
rules for both \texttt{aterm-eval} and \texttt{atermlist-eval}, it suffices
to invoke:
\begin{verbatim}
  (deffixequiv-mutual aterm-eval)
\end{verbatim}
The \texttt{deffixequiv-mutual} macro determines the types of the arguments by
examining the guards specified in the \texttt{define} formals, and it uses the
flag induction scheme produced by {\tt defines} to automatically prove the
congruence.

For recursive or mutually-recursive functions, proving a congruence directly
can be difficult because there are two calls of the function in the statement
of the theorem, and these two calls may suggest different induction schemes
that may not be simple to merge.  However, the congruences we are concerned
with follow from the fact that the fixing function is transparent to the
function, which can usually be proved straightforwardly by induction on the
function's own recursion scheme.  In practice, the \texttt{deffixequiv} and
\texttt{deffixequiv-mutual} utilities usually fully automate the derivation of
the congruence from the transparency theorem.

Even if we only need to write a \texttt{deffixequiv} or
\texttt{deffixequiv-mutual} form after each definition, this can be easy to
forget.  To further automate following the discipline, you can optionally
enable a \textit{post-define hook} that will automatically issue a suitable
\texttt{deffixequiv} command after each definition.  See the documentation for
\Xtopiclink{FTY____FIXEQUIV-HOOK}{fixequiv-hook} for details.




\section{Challenges and Future Work}
\label{sec:future-work}

The FTY library provides a robust implementation of a type system that
would feel familiar to users of strongly typed functional programming
languages such as Haskell or ML.  However, there are a few pitfalls in
their practical use, which we discuss below along with potential
solutions.

\subsection{Generic Functions}

The most common problem in working with the fixtype discipline is in
the use of generic functions such as \texttt{assoc}, \texttt{append},
and many others.  These functions are designed to work on objects
of nonspecific type, and therefore don't follow fixing conventions for
specific types.    One
can always apply appropriate fixing functions to the inputs of these
functions, so programming with them in a fixtype discipline isn't
hard.  However, applying this simple strategy to theorems will
often result in ineffective rewrite rules.

For example, suppose
\texttt{(bind-square-to-root key alist)} fixes \texttt{key} to type
\texttt{natp} and \texttt{alist} to type \texttt{nat-nat-alist-p}, and
we want to prove a theorem like the following:

\begin{verbatim}
 (equal (assoc k (bind-square-to-root k rest))
        (or (and (square-p k)
                 (cons k (nat-sqrt k)))
            (assoc k rest)))
\end{verbatim}

Presumably this isn't true without some type assumptions.  One way to
fix the theorem is to apply fixing functions everywhere that typed
variables are used in generic contexts:

\begin{verbatim}
 (equal (assoc (nfix k) (bind-square-to-root k rest))
        (or (and (square-p k)
                 (cons (nfix k) (nat-sqrt k)))
            (assoc (nfix k) (nat-nat-alist-fix rest))))
\end{verbatim}

But this rewrite rule is not always applicable.  The left hand side
will match only when we have an explicit \texttt{nfix} in our goal,
but this \texttt{nfix} is likely to be simplfied away in cases where
the key is known to be a natural.
In these cases, a formulation with a type hypothesis would work:
\begin{verbatim}
 (implies (natp k)
          (equal (assoc k (bind-square-to-root k rest))
                 ...))
\end{verbatim}

Unfortunately, this rule typically won't allow us to simplify terms
such as:
\begin{verbatim}
  (assoc (nfix k) (bind-square-to-root k rest))
\end{verbatim}
because it fails to unify.
In general, both kinds of terms may be encountered and both
formulations of the rule may be useful.  To solve this problem, one
might consider automation to generate both forms of the theorem, or to
generate a theorem that catches both cases as follows:

\begin{verbatim}
  (implies (and (syntaxp (or (equal k1 k) (equal k1 `(nfix ,k))))
                (nat-equiv k1 k)
                (natp k1))
           (equal (assoc k1 (bind-square-to-root k rest))
                  ...))
\end{verbatim}

For the moment, unfortunately, reasoning about a mix of generic
functions with fixtype-discipline functions seems to require the sort
of consideration of types that we had hoped to avoid.  We generally
deal with these problems on an ad-hoc basis, either by proving both
forms of the theorem or, in more problematic cases, by introducing
typed alternatives to the generic functions involved.

\subsection{Subtypes}

It is possible to use the FTY library to define two types that have a
subtype relation, but the library doesn't have any automation for
proving or making use of this relationship. 




In practice, we have found it difficult to get subtype relations to
work well.  Proving theorems about a mixture of functions that operate
on sub- and supertypes has the same problems as proving theorems with
a mixture of generic and fixtype functions, as discussed above.
Reasoning about a subtype hierarchy also can lead to degraded prover
performance, since proving that something is of type $A$ may lead by
backchaining to attempting to prove it to be of each subtype of $A$.

\subsection{Parameterized Types}

Haskell and ML support types that take other types as parameters,
e.g., $\textsf{List}\ A$ signifying a list of objects of type $A$, where
$A$ is a type variable.  Function signatures may contain types
that are not fully specified, and these functions may later be used in
contexts where the type variables are concretized as particular types.

Selfridge and Smith \cite{14-selfridge-polymorphic} created a macro
library that supports a form of polymorphism by automating the
creation of instances of the \Xtopiclink{ACL2____DEFSUM}{defsum}
macro.  Polymorphic functions are then supported by another set of macros
that allow one to instantiate a template function definition with
different substitutions for type variables.  A similar macro library
could be used to add polymorphism via templates to FTY, but this
has not yet been done.

\subsection{Dependent Types}

Correct behavior of multiple-input functions often depends on
constraints involving more than one of the inputs.  The fixtypes
discipline is focused on unary types, but occasionally it is desirable
for a product type to contain multiple elements that have constraints
linking them.  We have experimentally implemented support for this in
\texttt{defprod} and \texttt{deftagsum} by allowing the user to
specify these constraints along with an extra fixing step that forces
the fields to satisfy these constraints; this works in practice for
simple constraints like ``a literal's value should fit into its
width.''  We expect that there would be difficulties in formulating
constraints between subfields of a recursive data structure.

\subsection{Symbolic and Logical Evaluation}

Execution efficiency of functions using the fixtype discipline is
highly dependent on the use of \texttt{mbe} to avoid calling fixing
functions.  Evaluation in the logic (with guard checking turned off)
is much more expensive with such functions because the \texttt{:logic}
part of the \texttt{mbe} then needs to be executed.

This problem also applies to symbolic evaluation with the GL system
\cite{11-swords-gl}.  GL is used in hardware verification at Centaur
and elsewhere; it evaluates ACL2 functions on bit-level symbolic
inputs, producing bit-level symbolic results, allowing the use of SAT
or BDD reasoning to prove ACL2 theorems.  However, GL ignores guards
(except for concrete evaluation) and instead symbolically simulates the logical
definitions of functions. Therefore, when using GL on
fixtype-compliant functions, these fixing functions will be
unnecessarily (symbolically) executed frequently.

This extra expense could be problematic in some cases.  In future work we
expect to address this by adding a facility to FTY to generate extra GL rules
to help it avoid executing fixing functions.  Currently, we have worked around
this problem in some cases by using fixing functions that are cheap to
symbolically execute.  For example, if we are dealing with, say, a 32-bit
unsigned integer type, then for symbolic simulation it is cheaper to use
\texttt{(loghead 32 x)} rather than \texttt{(if (unsigned-byte-p 32 x) x 0)} as
the fixing function.  This expense of fixing can also be avoided by creating
custom symbolic counterparts, which are used in important core routines in
hardware verification frameworks like \Xtopiclink{ACL2____ESIM}{ESIM} and
\Xtopiclink{ACL2____SV}{SV}.

\subsection{Traversal of Complicated Data Structures}

In languages like ML or Haskell, it is possible to write higher order
functions for traversing deeply nested data structures.  This
capability goes a long way toward making it reasonable to inspect and
manipulate such objects.  Since ACL2 is first order, we cannot write
these kinds of generic traversals.  Instead, we have to duplicate the
boilerplate code for traversing a structure in each algorithm that
operates on it.  This can become very tedious.

For example, the parsetree format for the VL Verilog/SystemVerilog
toolkit contains 168 datatypes, 132 of which are defined in terms of
other types (as a product, list, etc.), reflecting the complexity of
the SystemVerilog language.  We might like to, for instance, collect
all identifiers used in a module.  We might also like simplify all
expressions throughout a module.  Doing either of these will require
traversing many of the same structures (modules, declarations,
assignments, etc.) to reach the objects of interest (identifiers,
expressions).

We have implemented an experimental utility, \texttt{defvisitor},
intended to generate the boilerplate code necessary for these
situations.  The user provides code to be run on certain types and to
combine results from recursive calls, and the utility generates the
boilerplate to traverse the data structures.  The current
implementation is a proof of concept and its user interface is likely to
change, but it has been used to implement several algorithms within
the VL library.



\section{Related Work}
\label{sec:related-work}

\subsection{Fixing Conventions}

The use of fixing conventions to avoid hypotheses is well studied and has been
used since the earliest Boyer Moore provers.  In Boyer and Moore's 1979 \emph{A
  Computational Logic} we find a \texttt{fix} function for NQTHM's naturals and
hypothesis-free theorems such as the commutativity of plus.  Boyer and Moore
credit A. P. Morse as the inspiriation for this approach, citing his treatment
of set theory, \emph{A Theory of Sets}~\cite{65-morse-sets}, and
recalling\footnote{Correspondence with Bob Boyer and J Moore.} that:
\begin{quotation}
``Morse tried every way he could to fix every function that he introduced to
eliminate hypotheses if possible, without doing any damage.  He delighted in
such theorems as that \textit{and} and \textit{or} distributed over one another
for all arguments, no matter what objects the arguments were, no matter that
\textit{and} was the exact same as set intersection and that \textit{or} was
the same as set union.''
\end{quotation}
In their 1994 \emph{Design Goals for ACL2}, Kaufmann and Moore reflect that
``NQTHM's logic gets incredible mileage out of the notion that
functions---especially arithmetic functions---default `unexpected' inputs to
reasonable values so that many theorems are stated without hypotheses.''  As a
result, fixing conventions were used liberally in their new theorem prover,
e.g., throughout its completion axioms for primitive functions on numbers,
characters, strings, etc.

%
%
%

Since then, many ACL2 libraries such as
\texttt{std/osets}~\cite{04-davis-osets} and
\Xtopiclink{ACL2____BITOPS}{bitops}, have made heavy use of the technique.
Perhaps the most extreme examples are found in
\texttt{misc/records}~\cite{02-kaufmann-records} and related work such as typed
records~\cite{03-greve-typedrecords}, \Ldef{defexec}-enhanded
records~\cite{07-greve-efficient}, memories~\cite{06-davis-memories}, and
\Xtopiclink{ACL2____DEFRSTOBJ}{\Ldef{defrstobj}}, which each use sophisticated,
convoluted fixing functions to achieve hypothesis-free read-over-write
theorems.

Fixing conventions are often unnecessary in typed logics.  In such a logic,
when we define functions such as $ \textsc{plus} : \textsc{Nat} \times
\textsc{Nat} \rightarrow \textsc{Nat} $, there is no need to include any type
hypotheses in theorems such as the commutativity of \textsc{plus}, because any
attempt to call or reason about \textsc{plus} on non-\textsc{Nat} arguments is
simply an error.  On the other hand, even in such a logic, fixing conventions
may be useful for modeling behavior of operations whose intended domains are
not easy to describe using types.  Lamport and Paulson~\cite{99-lamport-typed}
provide an engaging discussion of these sorts of issues.

\subsection{Data Struture Libraries}

There has been significant previous work to develop data structure libraries
for ACL2.  An early example is Brock's classic
\Xtopiclink{ACL2____DATA-STRUCTURES}{data-structures} library, which featured
macros such as
\Xtopiclink{ACL2____DEFSTRUCTURE}{\Ldef{defstructure}}~\cite{97-brock-defstructure}.
As a concrete example of using this macro, we might write:

\begin{verbatim}
   (defstructure student
     (name (:assert (stringp name) :type-prescription))
     (age  (:assert (natp age) :type-prescription)))
\end{verbatim}

\noindent This produces a constructor that simply conses together its arguments
and accessors that simply \Lfn{car}/\Lfn{cdr} into their argument.  No fixing
is done, so the constructor only produces a well-formed \Lfn{student-p} if
its arguments have the proper types, and the accessors may produce ill-typed
results when applied to non-\Lfn{student-p} objects.  Accordingly, reasoning
about such structures typically requires type hypotheses.  The more recent
\Xtopiclink{STD____DEFAGGREGATE}{\Ldef{defaggregate}} macro follows this same
approach.

ACL2's single threaded objects~\cite{02-boyer-stobjs} are in many ways similar
to \texttt{defstructure} and \texttt{defaggregate}.  Although it probably would
make little sense to define a student structure as a stobj, we can do so:

\begin{verbatim}
(defstobj student
  (name :type string        :initially "")
  (age  :type (integer 0 *) :initially 0))
\end{verbatim}

\noindent The resulting recognizer, accessors, and mutators are similar to
those produced by \Ldef{defstructure} or \Ldef{defaggregate} and so reasoning
about these operations usually requires type hypotheses.  On the other hand,
the recent addition of abstract stobjs~\cite{13-goel-stobjs} makes it
possible to develop alternative logical interfaces, e.g., we could arrange so
that the \texttt{student} stobj was logically viewed as an FTY product object.

The \Xtopiclink{ACL2____DEFDATA}{defdata}~\cite{14-harsh-defdata}
library by Chamarthi, Dillinger, and Manolios features an alternative
macro, also called \Ldef{defdata}, that supports introducing richer
types, such as sum types, mutually recursive types, etc.  This
framework also features integrated support for counterexample
generation, an exciting feature which FTY does not yet have.  To
define a similar \texttt{student} structure with \Ldef{defdata} we
might write:




\begin{verbatim}
  (defdata student (record (name . string)
                           (age  . nat)))
\end{verbatim}

\noindent This similarly results in a \Lfn{studentp} recognizer,
constructor, and accessors like \Lfn{student-age}.  Unlike
\Ldef{defstructure} or \Ldef{defaggregate}, the underlying representation
of these functions is based on an optimized variant~\cite{07-greve-efficient}
of Kaufmann and Sumners' records book~\cite{02-kaufmann-records}, and the macro
provides special integration with ACL2's
\Xtopiclink{ACL2____INTRODUCTION-TO-THE-TAU-SYSTEM}{Tau} reasoning procedure.
However, the general approach to type reasoning about these structures is
unchanged: we still require type hypotheses to establish that the construct
produces a valid \Lfn{studentp}, that \Lfn{student-name} returns a
string, and so forth.

Despite type hypotheses, macros like \Ldef{defstructure}, \Ldef{defaggregate},
and \Ldef{defdata} are certainly very useful.  \Ldef{Defstructure} has long
been used in ACL2 developments, including recent work such as the modeling by
Hardin, et al.~\cite{14-hardin-llvm} of the LLVM compiler project's
intermediate form in ACL2, and the formalization by van Gastel and
Schmaltz~\cite{13-gastel-xmas} of the xMAS language for communication networks
on multi-core processors and systems-on-chip.  For many years, we used
\texttt{defaggregate} and other
\Xtextlink{ACL2____STD_F2UTIL}{\texttt{std/util}} macros at Centaur as the
basis for our \Xtextlink{ACL2____VL}{VL} library, microcode
model~\cite{14-centaur-ucode}, and other internal applications. In our
experience, porting these libraries to FTY was not difficult and has helped to
simplify our code.


\subsection{Make-Event Metaprogramming}

In 2004, Vernon Austel developed~\cite{04-austel-typing} an experimental
variant of ACL2 that added support for a certain type system.  He explained
that this work had required modifying ACL2 because ``a usable type system must
constantly extend the set of functions whose type it knows about; this seems to
require storing type information in the ACL2 world, which macros currently
cannot do,'' and proposed extending ACL2 with something like
\Xtopiclink{ACL2____MAKE-EVENT}{make-event} to ``allow others to experiment
with type systems without having to hack the system code.''  Indeed, our FTY
library makes extensive use of \texttt{make-event} to record the associations
between type recognizers, fixing functions, and equivalence relations, and to
look up (via \texttt{define}) the type signatures for functions.

\section{Conclusion}

FTY is a new data structure library for ACL2 that provides deep support for
using fixing functions to avoid type hypotheses in theorems.  Its successful
use may require somewhat more discipline than similar libraries such as
\texttt{std/util} or \texttt{defdata}.  In exchange, it provides a strongly
typed programming environment that can help to catch errors during development
while largely avoiding type hypotheses during theorem proving.

Having a good data structure library is tremendously useful when developing
large systems in ACL2.  A fixing discipline is one part of this, but FTY is
also increasingly mature and capable in other ways, e.g., it retains much of
the \texttt{std/util} look and feel, with features such as
\Xtopiclink{ACL2____XDOC}{XDOC} integration, convenient
\Xtopiclink{ACL2____B_A2}{b*} binders, readable \texttt{make}/\texttt{change}
macros, etc.  We are now using FTY as for large ACL2 libraries such as SV and
VL libraries, and have been pleased with the results.

The source code for FTY is included in the ACL2
\Xtextlink{ACL2____COMMUNITY-BOOKS}{Community Books} under the
\texttt{centaur/fty} directory.  Beyond this paper, the FTY library has
extensive \Xtextlink{ACL2____FTY}{documentation}, which includes more detailed
information on the available options for each macro.  The \texttt{centaur/fty}
directory also includes various test cases that may serve as useful examples of
using the library.

We hope you find FTY useful.

\subsection{Acknowledgments}

We thank Bob Boyer and J Moore for very interesting discussions about the
origins of fixing disciplines in Boyer-Moore provers.  We thank Shilpi Goel and
Cuong Chau for their corrections and feedback on this paper.

\bibliographystyle{eptcs}
\bibliography{paper}
\end{document}